# Investigating and Controlling the Libration and Rotation Dynamics of Nanoparticles in an Optomechanical System

Chaoxiong He[1], Jinchuan Wang[1], Ying Dong[1], Shaochong Zhu[2], Qianwen Ying[1], Yuanyuan Ma[5], Fu Feng[1], Zhangqi Yin[4], Cuihong Li[1,*] and Huizhu Hu[3]

[1]*Research Center for Frontier Fundamental Studies, Zhejiang Lab, Hangzhou 310000, China*

[2]*Research Center for Novel Computational Sensing and Intelligent Processing, Zhejiang Lab, Hangzhou 310000, China*

[3]*Laboratory of Extreme Photonics and Instrumentation, College of Optical Science and Engineering, Institute of Quantum Sensing, Zhejiang University, Hangzhou 310027, China*

[4]*Center for Quantum Technology Research and Key Laboratory of Advanced Optoelectronic Quantum Architecture and Measurements (MOE), School of Physics, Beijing Institute of Technology, Beijing 100081, China*

[5]*The State Key Lab of Analytical Chemistry for Life Science, School of Chemistry and Chemical Engineering, Chemistry and Biomedicine Innovation Center (ChemBIC), Nanjing University, Nanjing 210000, China*

*\*e-mail: licuihong@zhejianglab.com*

In optomechanical systems, the libration and rotation of nanoparticles offer profound insights for ultrasensitive torque measurement and macroscopic quantum superpositions. Achievements include transitioning libration to rotation up to 6 GHz and cooling libration to millikelvin temperatures. It is undoubted that the libration and rotation are respectively driven by restoring and constant optical torques. The transition mechanisms between these two states, however, demand further exploration. In this perspective, it is demonstrated in this manuscript that monitoring lateral-scattered light allows real-time observation of libration/rotation transitions and associated hysteresis as ellipticities of trapping laser fields vary. By calculating optical torques and solving the Langevin equation, transitions are linked to the balance between anisotropic-polarization-induced sinusoidal optical torques and constant ones, with absorption identified as the main contributor to constant torques. These findings enable direct weak torque sensing and precise nanoparticle control in rotational degrees, paving the way for studying quantum effects like nonadiabatic phase shifts and macroscopic quantum superpositions, thereby enriching quantum optomechanics research.

Optically-levitated nanoparticles, acting as high-fidelity nano-oscillators and benefiting from their isolation from external environment, have emerged as superior facilitators for probing nanoscale photonic and material interactions[1-3], exploratory tools for delineating the quantum-classical boundary[4-6], and sensors for ultraweak forces[7-9]. The investigation into their center-of-mass translational oscillations, along with inherent nonlinear dynamics induced by optical forces, has been extensive and profound[10-13]. Furthermore, optically-levitated nanoparticles exhibit a spectrum of motions including librations, rotations, and spins. This rotational dynamism endows them with heightened sensitivity for measurements pertaining to rotation[14-19], positioning them as quintessential platforms for the study of advanced mesoscopic phenomena. Notable among these are the exploration of vacuum friction[15], the manifestation macroscopic quantum coherence[20], the interrogation of spin-mass interaction[21], the analysis of many-body correlations[22], and the examination of quantum nonadiabatic phenomena[23], thereby significantly contributing to the advancement of mesoscopic physics and quantum science.

It has been elucidated that the major axis of non-spherical nanoparticles exhibits a tendency to librate around the polarization vector of the trapping laser, a phenomenon attributed to the anisotropic polarization and librational balance[24-26]. For non-spherical nanoparticles trapped by circularly polarized laser fields, they are subjected to rotational motions, frequencies of which are governed by the equilibrium established between constant optical torques and viscous drags[25,27-29]. Noteworthy achievements include the attainment of libration cooling to temperatures in the millikelvin range[30-32] and the realization of rapid mechanical rotations of nanoparticles, having surpassed 6 GHz[28]. A recent investigation has delved into the rotational diffusion phenomenon, exploring the noise-induced transition between the libration and rotation of an optically levitated nanodumbbell[33]. Despite these advances, the transition between librations and rotations remains sparsely documented, underscoring a significant gap in understanding the fundamental mechanisms that underpin mesoscopic dynamics and quantum spin-mechanical interactions in optomechanical systems[21,34,35].

This article proposes a novel methodology for monitoring the librational and rotational dynamics of nanoparticles by analyzing lateral-scattered light as ellipticities of trapping laser fields are varied under diverse trapping powers and vacuum degrees. This approach facilitates the direct observation of real-time critical librational/rotational transitions of nanoparticles, which are characterized by stochastic exchanges between librations and rotations. Moreover, these transitions are marked by distinctive bi-stable hysteresis loops in relation to ellipticities. To account for transition mechanisms, optical torques, quantified as integrals over angular momentum current density, are computed, enabling the resolution of librational/rotational Langevin equations. The calculation results show perfect agreements with experiment. Further theoretical examination indicates that, the critical libration/rotation transitions in elliptically polarized laser fields primarily arise from the balance between anisotropic-polarization-induced sinusoidal optical torques and constant ones, and absorption processes play a dominant role in generating constant optical torques. These findings establish a foundational understanding for the precise control of nanoparticles' dynamics, paving the way for advancements in the exploration of mesoscopic dynamics and macroscopic quantum effects within the realm of optomechanics.

As is shown in Fig.1 (a), our experimental setup includes an ellipticity-controllable 1064 nm trapping laser field and a linearly-polarized 532 nm detecting laser field. By rotating the half-wave plate before several high-reflective mirrors, we manage to simultaneously control ellipticities $\eta$ and polarization azimuth angles $\theta_0$ of the 1064 nm trapping laser field. Both laser fields propagate along the $z$ axis, and the 532 nm laser is linearly polarized on the $x$ axis. $\theta$ ($\varphi$) represent the angle between the $x$ ($z$) axis and the projection of the nanodumbbell's symmetrical axis in the $xy$ ($xz$) plane. The lateral-scattered 532 nm power is measured to rapidly confirm the nanodumbbell's geometries[36,37]. The 1064 nm light forward-scattered from the nanodumbbell and the forward-propagating trapping laser beam, interfered with each other, which makes it feasible to detect the nanodumbbell's motions with far-field photodetectors[30,38].

As shown in Fig.1 (b)-(d), distinct features can be observed in librational/rotational Power Spectral Densities (PSDs) under ellipticities $\eta$ = 0.0, 0.8 and 1.0. Only libration, indicated by the narrow spectrum around 300 kHz, can exist under the linearly-polarized trapping laser field ($\eta$ = 0.0), while only rotation, indicated by wide spectrums, can exist under the circularly polarized trapping laser field ($\eta$ = 1.0). In particular, when $\eta$ = 0.8, the corresponding PSD combines both librational and rotational spectrums, from which we can infer that the nanodumbbell is probable to undergo both libration and rotation under this circumstance.

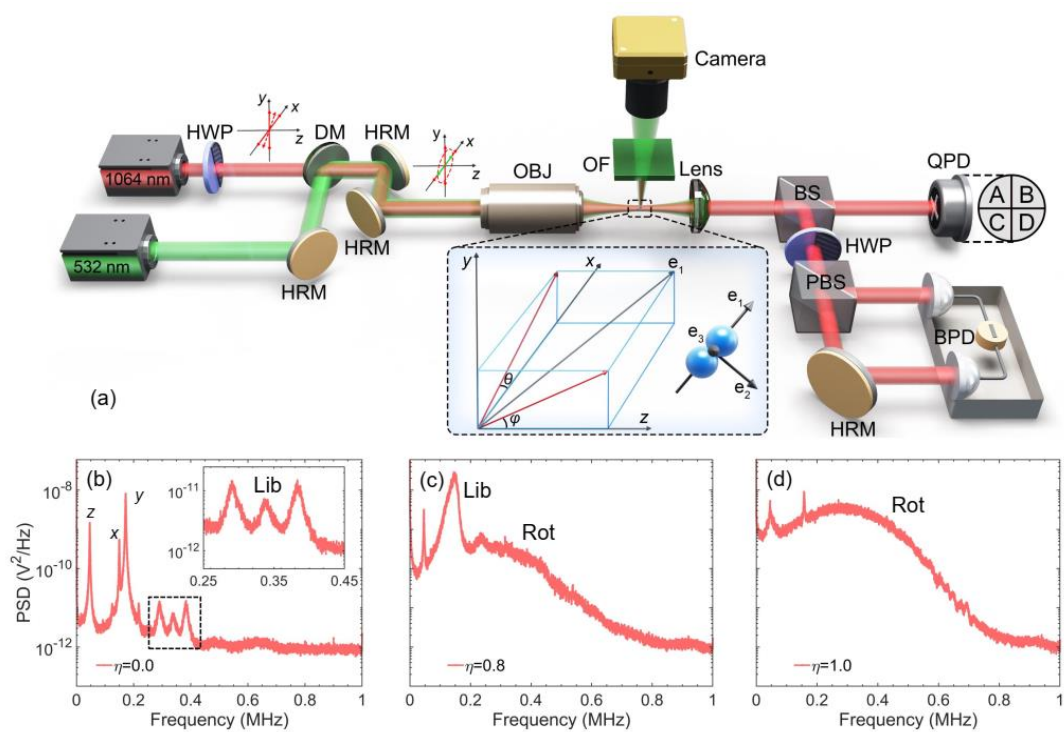

**Fig. 1 Experimental Overview.** (a) Experimental setup and definitions of angular coordinates. HWP: Half Wave Plate; DM: Dichromic Mirror; HRM: High-Reflective Mirror; OBJ: Objective Lens; OF: Optical Filter; BS: Beam Splitter; PBS: Polarization Beam Splitter; QPD: Quadrant Photodetector; BPD: Balanced Photodetector. (b) A librational PSD when $\eta$ = 0.0. (c) A PSD revealing both the libration and the rotation when $\eta$ = 0.8. (d) A rotational PSD when $\eta$ = 1.0. (b), (c) and (d) are measured under the vacuum degree of around 2 mbar and the trapping laser's power of around 100 mW. Lib and Rot represent the librations and rotations respectively.

To get a more particular knowledge of the librational/rotational dynamics under the manipulation of ellipticities, we measure librational/rotational PSDs and lateral-scattered 532 nm powers while changing trapping laser fields' polarizations. As shown in Fig.2(a), when the angles of the half-wave plate $\theta_{\lambda/2}$ increase from 0° to approximately 50°, the 1064 nm trapping laser field undergoes a transformation where its polarization azimuth angles $\theta_0$ nonlinearly increase from 0° to nearly 90°, and simultaneously, its ellipticities $\eta$ vary from zero to a maximum and then back to zero. The ellipticity $\eta$ reaches the maximum of about 0.8 when the polarization azimuth angle $\theta_0$ approaches 45°. Inspired by the fact that lateral-scattered 532 nm powers can directly reflect the alignment of the nanodumbbell[37], we extend lateral-scattered 532 nm powers to observe critical librational/rotational transitions in real time. As shown in Fig.2(b), referring to PSDs, the maxima and the minima of 532 nm scattered powers reveal the nanodumbbell's steady librations around $x$ and $y$ axes respectively, while intermediate values indicate the nanodumbbell's continuous rotations about the $z$ axis. When the polarization azimuth angle $\theta_0$ of the 1064 nm trapping laser deviates slightly from the $x$ (or $y$) axes and accompanied by moderate values of the ellipticity $\eta$, Fig.2(c) depicts that lateral-scattered 532 nm powers exhibit stochastic interchange between the intermediate values and the maxima (or minima). The observed interchange signifies real-time librational/rotational transitions, and we define the corresponding ellipticities as "critical ellipticities". Therefore, compared with PSDs, lateral-scattered 532 nm powers capture the librational/rotational dynamics in a rapid, effective and intuitive way.

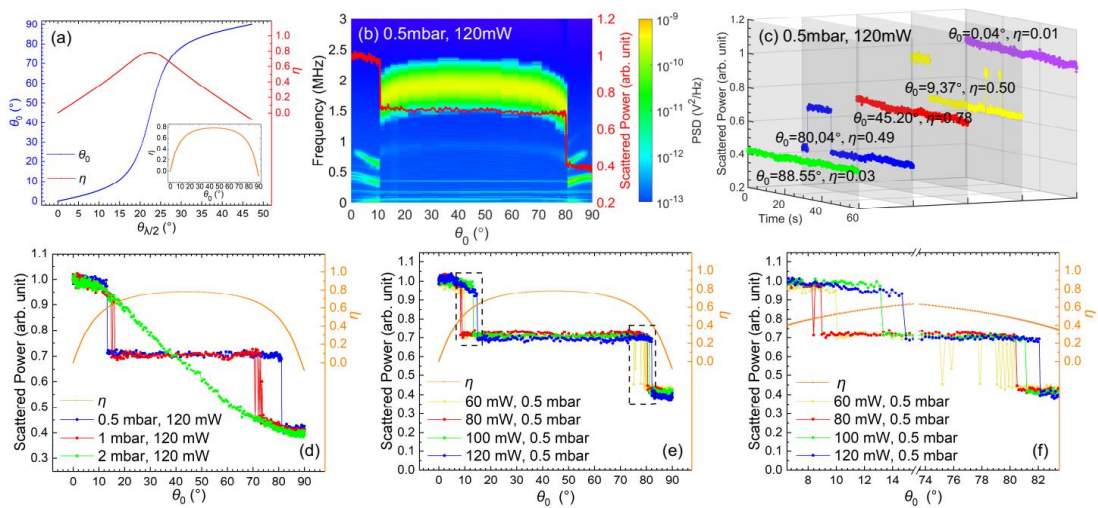

**Fig. 2 Experimental results of polarization control and librational/rotational manipulations.** (a) Variations of ellipticities $\eta$ and polarization azimuth angles $\theta_0$ of 1064 nm trapping laser fields

as the half-wave plate rotates. Inset: ellipticities $\eta$ as a function of polarization azimuth angles $\theta_0$. (b) Referring to PSDs, lateral-scattered 532 nm powers are reliable criteria to extract the librational/rotational dynamics. The maxima and minima of scattered 532 nm powers reveal librations around $x$ and $y$ axes respectively while the intermediate values reveal rotations. (c) Visualized observations on librations, rotations and librational/rotational transitions with lateral-scattered 532 nm powers. (d) Critical librational/rotational transitions under different vacuum degrees. (e) Critical librational/rotational transitions under different trapping laser powers. (f) Zoom-in of dashed boxes in (e).

"Critical ellipticities" are crucial for explorations and manipulations on the mesoscopic dynamics of librational/rotational motions. The critical ellipticities subjected to four typical different vacuum degrees are shown in Fig. 2(d). A reduction in critical ellipticities correlates with the decrease in gas pressures. The observation can be qualitatively explained with damping torques from ambient gas molecules. Higher gas pressures will generate stronger damping torques, and thus hinder nanodumbbells' transitions from librations to rotations by restraining their librations' angular amplitudes, or else facilitate their transitions from rotations to librations by decelerating rotations. The effect of trapping laser powers on librational/rotational transitions is shown in Fig.2(e) and (f). It is obvious that the critical librational/rotational transition behaviors become more noticeable as trapping laser powers increase. Intriguingly, Fig.2(e) and (f) highlight a significant contrast in critical ellipticities between libration-to-rotation and rotation-to-libration transitions.

To deeply visualize the contrast above, we tune the polarization azimuth angles in two opposite directions with the trapping laser power fixed to about 120 mW and the gas pressure around 0.5 mbar. As a result, Fig.3 shows the noticeable hysteresis-loop structure of critical librational/rotational transitions with respect to ellipticities. The hysteresis loop proves the bi-stability of critical librational/rotational transitions, which highlights the significant impact of the nanodumbbell's initial states in critical librational/rotational transitions.

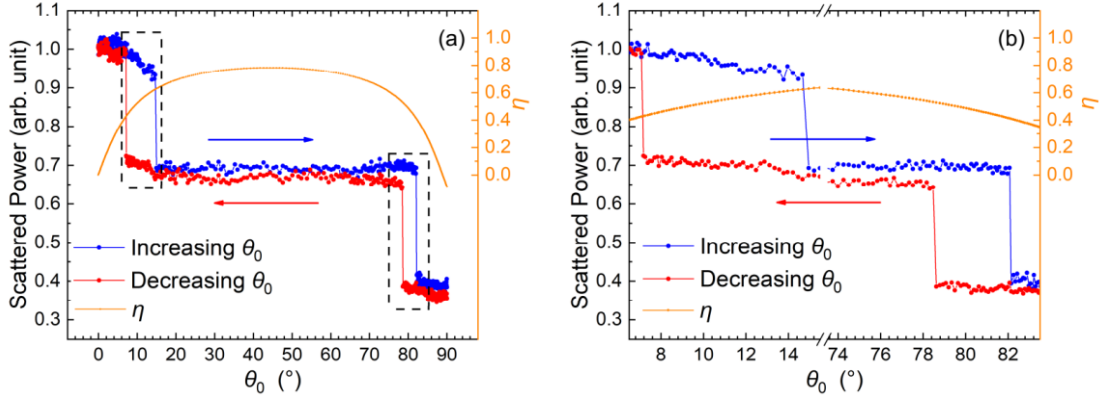

**Fig. 3 Experimental results of bi-stable hysteresis-loop structures of critical librational/rotational transitions.** (a) Lateral-scattered 532 nm powers are measured in response to tuning polarization azimuth angles $\theta_0$ of 1064 nm trapping laser fields at two opposite directions under the vacuum degree around 0.5 mbar and the trapping laser power around 120 mW. Arrows represent tuning directions of polarization azimuth angles $\theta_0$. (b) Zoom-in of dashed boxes in (a).

In order to provide quantitative illustrations for bi-stabilities in critical librational/rotational-transition dynamics, we carry out ab-initio calculations on optical torques $\boldsymbol{\tau}_{opt}$ according to the angular-momentum-conservation law[39-41] (See Methods for details) and subsequently numerically solve the librational/rotational Langevin equation(1):

$$I\frac{d^2\boldsymbol{\theta}}{dt^2} = -I\Gamma\frac{d\boldsymbol{\theta}}{dt} + \boldsymbol{\tau}_{opt} + \boldsymbol{\tau}_{thermal} \quad (1)$$

where $I$ and $\Gamma$ stand for rotational inertias and damping rates, and $\boldsymbol{\tau}_{thermal}$ represents thermal-noise torques that can be expressed by a Gaussian stochastic process: $\langle \tau_{thermal}(t) \rangle = 0$, $\langle \tau_{thermal}(t)\tau_{thermal}(t') \rangle = 2Ik_B T\delta(t-t')$. During the calculations, only libration/rotational motions around the $z$ axis are considered and ellipticities of the optical fields are quasi-continuously increased or decreased. Calculated librational/rotational PSDs at the estimated vacuum around 0.5 mbar are shown in Fig. 4. The SiO$_2$ nanoparticle's complex refractive index is set to $\tilde{n} = 1.45875 + 0.00098i$ [42]. As is shown in Fig.4, under the laser power of 120 mW, there is an evident difference between critical ellipticities for libration-to-rotation and rotation-to-libration transitions, and under the laser power of 60 mW, the critical behavior and bi-stability become less obvious. Calculated results of bi-stabilities in Fig.4 perfectly agree with the experimental results shown in Fig.2 and Fig.3.

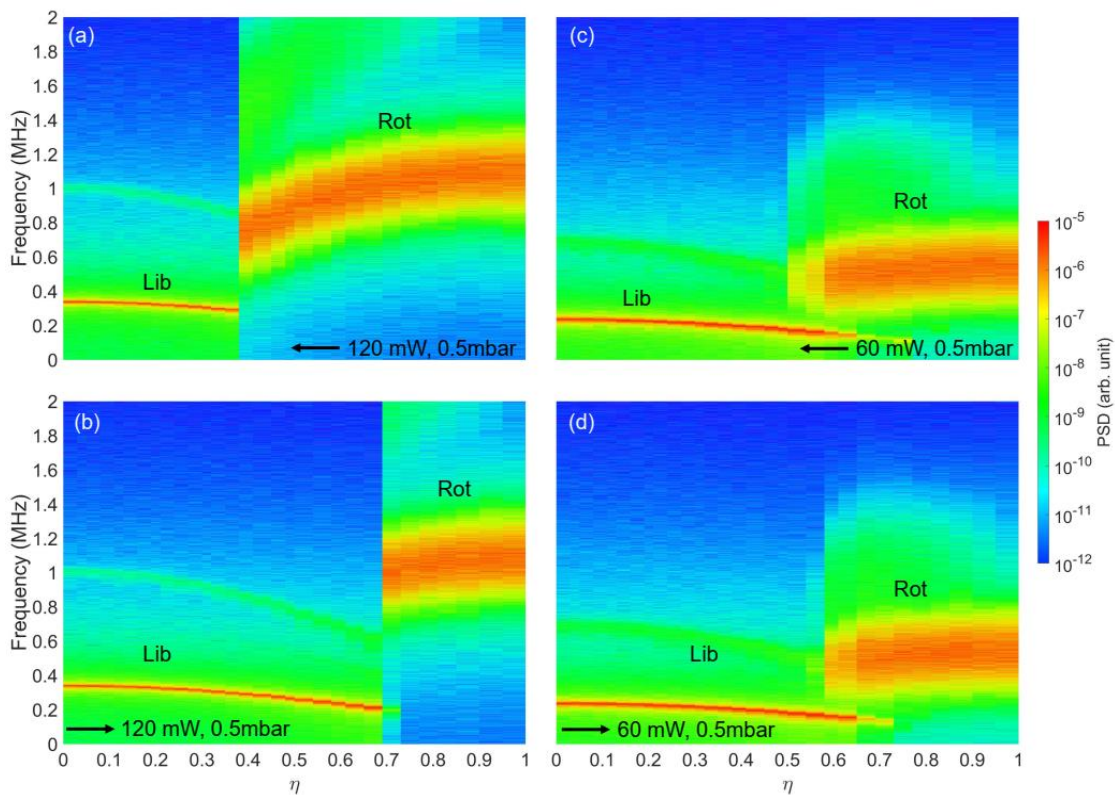

**Fig. 4 Calculation results of librational/rotational bi-stabilities.** During the calculation, ellipticities $\eta$ are quasi-continuously decreased or increased, and arrows stand for tuning directions of ellipticities. (a)-(b) Calculated librational/rotational PSDs under the fixed trapping laser power of 120 mW and the vacuum degree of 0.5 mbar. (c)-(d) Calculated librational/rotational PSDs under the trapping laser power of 60 mW and the vacuum degree of 0.5 mbar.

To investigate physical mechanisms behind the bi-stability with respect to ellipticities, it is required to analyze how optical torques are arisen. The widely-accepted dipole model divides optical forces (or torques) into two contributions named "gradient" and "scattering" ones[33,43,44]. It is uncontroversial that "gradient" forces (or torques) are determined by derivatives of conservative potentials arisen from polarizations. "Scattering" forces (or torques) involve complex mechanisms such as radiation reactions[33,43], absorptions, etc.

For driving steady librations of nanoparticles, anisotropic-polarization-induced sinusoidal optical torques $\boldsymbol{\tau}_{ani}$ serve as restoring torques[18,30], which corresponds with abovementioned "gradient" torques. Particularly, for the nanodumbbell's librations in the *xy* plane around the *z* axis, $\tau_{ani,z} = -\tau_0 \sin[2(\theta - \theta_0)]$, where $\theta_0$ equals to the polarization azimuth angle of the trapping laser in the *xy* plane, and $\tau_0$ is proportional to the laser power and jointly determined by the laser field's ellipticity and the nanoparticles' anisotropic parameters. (See Methods for the theoretical model in detail).

Referring to driving continuous rotations of nanoparticles, previous studies considered radiation reactions as the sole cause to exert constant scattering torques[33,45]. However, owing to optical absorptions, the spin angular momenta carried by absorbed photons will also exert constant torques $\boldsymbol{\tau}_{abs}$ on nanoparticles[46,47]. Here, to reveal the role of absorptions in constant optical torques, we numerically calculate the biased constant torques when the imaginary part $\kappa$ of the nanoparticle's complex refractive index $\tilde{n}$ changes from 0 to 0.005 with a step of 0.0002 and the real part *n* is fixed at 1.45875. During the calculation, the power and ellipticity of the 1064nm trapping laser field are fixed at 100mW and 0.69 respectively. Calculated results are shown in Fig. 5. In Fig.5, it is evident that as the imaginary part of the complex refractive index increases, the absorption becomes increasingly significant. When the imaginary part $\kappa$ of complex refractive index is zero, the biased constant optical torque remains at a magnitude of $10^{-22}$ N·m which is attributed to the radiation reaction[43] and consistent with results reported in Reference[33]. The complex refractive index utilized in Fig.4 is highlighted. It can be seen that the highlighted biased constant torque is on the magnitude of $10^{-21}$ N·m , which is one order of magnitude larger than that solely induced by the radiation reaction. Therefore, our calculations emphasize the dominant and non-negligible roles of absorptions to exert constant torques that drive continuous rotations of nanoparticles.

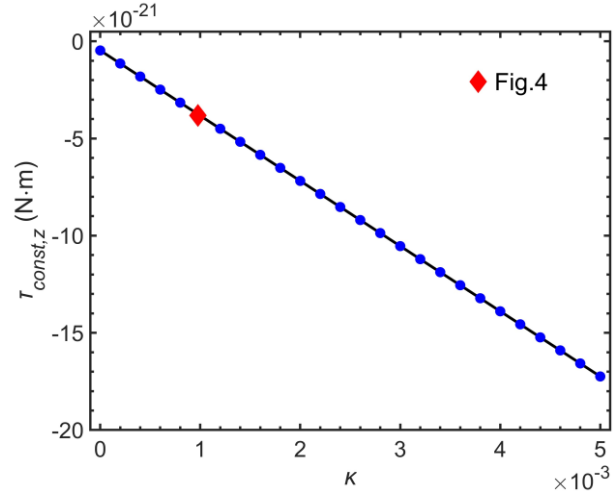

**Fig 5. Calculated biased constant torques as functions of $\kappa$.** $\kappa$ represent the imaginary part of the nanoparticle's complex refractive index. Calculations are done under the trapping laser power of 100 mW and the ellipticity of 0.69. The complex refractive index utilized in Fig.4[42] is highlighted. As $\kappa$ increases, the absorption becomes increasingly dominant in inducing the constant torque.

Based on the optical torques $\tau_{opt}$ above, the critical and bi-stable behaviours in librational/rotational transitions are straightforward to depict in the framework of washboard potential[48,49]. Whether the nanodumbbell experiences steady librations, continuous rotations or critical librational/rotational transitions, is mainly determined by the barrier-trap potential difference, its initial librational/rotational kinetic energy and stochastic damping. When the ellipticity is relatively small, the barrier-trap potential difference is relatively high and sinusoidal torques $\tau_{ani}$ play dominant roles, and the nanodumbbell tends to undergo librations. When the ellipticity is relatively large, the barrier-trap potential difference is relatively small, or else barriers and traps may even vanish, and constant torques $\tau_{const}$ become dominant and the nanodumbbell usually experiences rotations. At certain ellipticities with moderate barrier-trap potential difference, the significant stochastic damping from ambient gas molecules accounts for critical librational/rotational transitions. The balance between $\tau_{ani}$ and $\tau_{const}$ from elliptically-polarized laser fields constitutes physical basis for manipulating nanodumbbells' librational/rotational transitions. Furthermore, the bi-stability arises essentially due to the effect of the initial librational/rotational kinetic energy in librational/rotational transitions. (See Supplementary Section2 for more details)

In conclusion, the librational and rotational dynamics of an optically-levitated nanodumbbell levitated in elliptically-polarized laser fields are studied. With lateral-scattered 532 nm powers revealing real-time behaviors of librational/rotational dynamics, we report the bi-stabilities of critical librational/rotational transitions with respect to trapping laser fields' ellipticities. Such transitions are attributed to occur on account of the balance between anisotropic-polarization-induced sinusoidal optical torques and constant ones that are dominantly induced by absorptions. Significant stochastic collisions from gas molecules account for the dependence of critical ellipticities on vacuum degrees. Within the framework of washboard potentials and taking nanoparticles' initial librational/rotational kinetic energy into account, hysteresis-loop bi-stabilities are straightforwardly understood. Our observations establish essential foundations for developing quantum precision measurement techniques[1,13,50] and exploring inherent frontier physical problems on mesoscopic laser-matter interactions[6,21,51] in the rotational degree of freedom.

## Methods

### Experimental Details

A 1064 nm trapping laser field (Precilasers, YDFL-1064-SF-10-CW) and a linearly-polarized 532 nm detecting laser field (LIGHTHOUSE, Spout-D-5W) are applied in our experimental setup. These two laser fields propagate parallelly with each other and are tightly focused by an OBJ lens (Nikon, MUE21900, NA=0.8). A silica nanodumbbell is trapped in the 1064 nm focal region inside a vacuum chamber. We install a camera (DataRay, S-WCD-LCM-C) directly over the nanodumbbell to measure its lateral scattered 532 nm power. The forward-propagated 1064 nm light is collected by a high-precision aspherical lens (Thorlabs, AL1512-C, NA=0.55) and then is split to two parts. One part is detected by a custom-built QPD to monitor the nanodumbbell's translational displacements. The other part successively goes through a 1064 nm HWP (Thorlabs, WPH10M-1064) and a PBS (Thorlabs, CCM1-PBS253/M), and is finally sent to a BPD (Thorlabs, PDB450C-AC) to extract the nanodumbbell's librational/rotational motions.

The power of the 532 nm detection laser is deliberately maintained at a substantially lower level (~10 mW) compared to the 1064 nm trapping laser (~60-120 mW) to ensure that the control exerted by the 1064 nm trapping laser over the state of the trapped particle is not compromised. The 532 nm laser field is kept linearly polarized along the $x$ axis. A 1064 nm HWP is installed on a stepper motor rotation mount (Thorlabs, K10CR1/M) before several HRMs. HRMs are coated with dielectric films. By rotating the HWP, it is achievable to simultaneously control ellipticities $\eta$ and polarization azimuth angles $\theta_0$ of the 1064 nm trapping laser. Polarizations of both beams are measured with a compact polarimeter (Thorlabs, PAX1000) before conducting the experiments.

### Nanoparticle Trapping

Nanoparticles were loaded into the optical trap by using an ultrasonic nebulizer (Omron, NE-C25S). Nanoparticles were trapped at atmospheric pressure. A nanodumbbell is constituted with two adhered nanospheres (NanoCym, monodisperse silica nanoparticles with the nominal diameter of 150 nm). To identify a nanodumbbell, two methods are usually applied. One refers to the ratio between translational damping rates along $y$ and $x$ axes in a trapping laser field linearly polarized on the $x$ axis, $\Gamma_y/\Gamma_x$, and specifically speaking, for a nanodumbbell, $\Gamma_y/\Gamma_x$ approaches 1.3. The other involves variations of lateral-scattered 532 nm powers in response to changes of polarization azimuth angles $\theta_0$ of 1064 nm trapping laser fields. When $\theta_0$ changes from 0° to 90°, for a nanodumbbell, lateral-scattered 532 nm powers undergoes a changing ratio of approximately 65%.

**Data Acquiring and Processing**

The motional PSDs of the nanodumbbell are extracted with the forward-propagated 1064 nm light. The light contains two contributions, the 1064 nm light forward-scattered from the nanodumbbell and the forward-propagating trapping laser beam. The interference between them constitutes the fundament for extracting the nanodumbbell's motions. Referring to a dipole-moment scattering model, the detected signal from the QPD is directly proportional to the nanodumbbell's translational displacement, and the signal from the BPD is proportional to $\sin 2\theta \sin^2 \varphi$. The detected signals are recorded with the Data Acquiring Module inside a Phase-locked Amplifier (Zurich Instruments, MFLI). Processing the detected signals with Fourier transform, motional PSDs are thus measured.

The lateral-scattered 532 nm powers, which directly reveal real-time behaviors of librational/rotational dynamics of the nanodumbbell, are simultaneously acquired with angles of the HWP $\theta_{\lambda/2}$. This part of experimental data is obtained with programs developed by ourselves.

**Ab-initio Numerical Calculations of Optical Torques**

The specific calculating steps are as follows.

Step1: calculating the tightly-focused trapping optical field

The tightly-focused trapping optical field is calculated at first, serving as the background optical field to extract the laser-nanoparticle interaction.

A collimated elliptically-polarized optical laser beam can be expressed as,

$$\mathbf{E}_{\text{collimated}} = \left(E_{x0}\hat{\mathbf{e}}_{\mathbf{x}} + E_{y0}\hat{\mathbf{e}}_{\mathbf{y}}\right) e^{-(x^2+y^2)/w_0^2} \quad (1)$$

where $w_0$ stands for the radius of the beam, $E_{x0}$, $E_{y0}$ stand for complex amplitudes of optical fields along the $x$ and $y$ directions respectively. When it is tightly focused by a focal lens with high numerical aperture $NA$, the optical field near the focus is given by Equation (2) according to the vector diffraction reported by Richards and wolf[52,53],

$$\mathbf{E}_{\text{focus}}(r,\theta,z) = \frac{ikfe^{-ikf}}{2} \begin{pmatrix} E_{x0}\left[I_{00} + I_{02}\cos 2\theta\right] + E_{y0}I_{02}\sin 2\theta \\ E_{x0}I_{02}\sin 2\theta + E_{y0}\left[I_{00} - I_{02}\cos 2\theta\right] \\ -2iE_{x0}I_{01}\cos\theta - 2iE_{y0}I_{01}\sin\theta \end{pmatrix} \quad (2)$$

where $r = \sqrt{x^2+y^2}$, $\theta = \arctan\frac{x}{y}$, and $f$ is the focal length of the focal lens, and,

$$\begin{cases} I_{00}(r,z) = \int_0^{\varphi_{\max}} e^{-f^2\sin^2\varphi/w_0^2}(\cos\varphi)^{1/2}\sin\varphi(1+\cos\varphi)J_0(kr\sin\varphi)e^{ikz\cos\varphi}d\varphi \\ I_{01}(r,z) = \int_0^{\varphi_{\max}} e^{-f^2\sin^2\varphi/w_0^2}(\cos\varphi)^{1/2}\sin^2\varphi J_1(kr\sin\varphi)e^{ikz\cos\varphi}d\varphi \\ I_{02}(r,z) = \int_0^{\varphi_{\max}} e^{-f^2\sin^2\varphi/w_0^2}(\cos\varphi)^{1/2}\sin\varphi(1-\cos\varphi)J_2(kr\sin\varphi)e^{ikz\cos\varphi}d\varphi \end{cases} \quad (3)$$

where $\varphi_{\max} = \arcsin NA$, $J_i$ stands for the i-th order Bessel Function.

Step2: solving Maxwell Equations

To analyze the optical scattering from the nanoparticle, the whole optical field $\mathbf{E}$ are usually separated into two parts: the background field $\mathbf{E}_{\text{bg}}$ and the scattering field $\mathbf{E}_{\text{scat}}$, and thus $\mathbf{E} = \mathbf{E}_{\text{bg}} + \mathbf{E}_{\text{scat}}$. Concretely, the tightly-focused elliptically-polarized optical fields in Equation (2) are background fields, $\mathbf{E}_{\text{bg}} = \mathbf{E}_{\text{focus}}$, and the whole optical field $\mathbf{E}$ fulfills Maxwell equations. In this work, Maxwell equations are numerically solved in a micrometer space with the finite element method.

Step3: calculating optical torques

Through the abovementioned two steps, we get both the electric field $\mathbf{E}$ and the magnetic field $\mathbf{H}$ of optical fields. Therefore, integrating the angular momentum current density of electromagnetic fields over a closed surface surrounding the naondumbbell $S$, we can the optical torques $\boldsymbol{\tau}_{opt}$ shown in Equation (4).

$$\boldsymbol{\tau}_{opt} = \oiint_S \mathbf{r} \times \ddot{\mathbf{T}} \cdot \mathbf{n} dS \quad (4)$$

where $\mathbf{n}$ represents corresponding unit normal vector, and $\ddot{\mathbf{T}}$ is Maxwell Tensor,

$$\ddot{\mathbf{T}} = \frac{1}{2} \text{Re} \left[ \varepsilon \mathbf{E} \mathbf{E}^* + \mu \mathbf{H} \mathbf{H}^* - \frac{1}{2} \left( \varepsilon \mathbf{E} \cdot \mathbf{E}^* + \mu \mathbf{H} \cdot \mathbf{H}^* \right) \ddot{\mathbf{I}} \right] \quad (5)$$

where $\varepsilon$, $\mu$ and $\ddot{\mathbf{I}}$ stand for dielectric constant, permeability and unit tensor respectively.

**Theoretical Model**

In the maintext, it has been mentioned that optical torques $\boldsymbol{\tau}_{opt}$ are composed of two dominant contributions $\boldsymbol{\tau}_{opt} = \boldsymbol{\tau}_{ani} + \boldsymbol{\tau}_{abs}$. Insightfully, from the perspective of laser-nanoparticle interactions, anisotropic polarizations and absorptions are two dominant mechanisms to achieve the angular momentum transfer. Anisotropic polarizations essentially involve the exchange of angular momenta between optical fields and nanoparticles, while absorptions denote unidirectional angular momentum transfer from optical fields to nanoparticles. The former causes changes in the macroscopic polarization states of optical fields, while the latter results in energy loss of optical fields. Anisotropic polarizations are mainly arisen from birefringence and anisotropic geometrical dimensions. The former is mainly investigated on crystal particles, while the latter is prominent on non-spherical particles such as dumbbells, rods, and ellipsoids. Ignoring the spin of the nanodumbbell's symmetrical axis, $\boldsymbol{\tau}_{ani}$ is determined by total conservative optical potential $U_{ani}$,

$$\boldsymbol{\tau}_{ani} = -\frac{\partial U_{ani}}{\partial \varphi}\mathbf{e_y} - \frac{\partial U_{ani}}{\partial \theta}\mathbf{e_z} \quad (6)$$

where $U_{ani} = -\frac{1}{2}\langle \mathbf{p}_{ani} \cdot \mathbf{E}_{inc}\rangle$, $\mathbf{p}_{ani} = \ddot{\mathbf{R}}^{-1}\ddot{\boldsymbol{\alpha}}\ddot{\mathbf{R}}\mathbf{E}_{inc}$ represents the induced dipole moment, $\mathbf{E}_{inc} = (E_x\hat{\mathbf{e}}_\mathbf{x} + E_y\hat{\mathbf{e}}_\mathbf{y} + E_z\hat{\mathbf{e}}_\mathbf{z})e^{j\omega t}$ represents the tightly-focused trapping optical field, $\ddot{\boldsymbol{\alpha}} = diag(\alpha_x, \alpha_y, \alpha_z)$ stands for the polarizability tensor in the particle frame, and $\ddot{\mathbf{R}}$ stands for the rotation matrix to transform the particle frame to the laboratory frame.

$$\ddot{\mathbf{R}} = \begin{pmatrix} \cos\varphi & 0 & -\sin\varphi \\ 0 & 1 & 0 \\ \sin\varphi & 0 & \cos\varphi \end{pmatrix} \begin{pmatrix} \cos\theta & \sin\theta & 0 \\ -\sin\theta & \cos\theta & 0 \\ 0 & 0 & 1 \end{pmatrix} \quad (7)$$

In our mathematical derivation, we elucidate components of $\boldsymbol{\tau}_{ani}$ motivating torsions/rotations in the *xz* plane around the *y* axis and in the *xy* plane around the *z* axis, and results are shown in Equation (8),

$$\begin{aligned}\tau_{ani,y} &= -\frac{1}{4}(\alpha_x - \alpha_z)\sqrt{(|E_x|^2 - |E_z|^2)^2 + 4(E_x^*E_z + E_xE_z^*)^2}\cos(2\varphi - \varphi_0) \\ \tau_{ani,z} &= \frac{1}{4}(\alpha_y - \alpha_z)\sqrt{(E_x^2 + E_y^2)(E_x^{*2} + E_y^{*2})}\sin[2(\theta - \theta_0)]\end{aligned} \quad (8)$$

where $\varphi_0 = \arctan\frac{|E_x|^2 - |E_z|^2}{2(E_x^*E_z + E_xE_z^*)}$ approximates $\pi/2$ based on the assumption $|E_x| \Box |E_z|$ and $\theta_0 = \frac{1}{2}\arctan\frac{E_x^*E_y + E_xE_y^*}{|E_x|^2 - |E_y|^2}$ represents the polarization azimuth angle of the trapping laser fields. Therefore, the conservative components of $\boldsymbol{\tau}_{ani}$ show sinusoidal dependences on the nanodumbbell's angular displacements, and the amplitudes of $\boldsymbol{\tau}_{ani}$ decrease with ellipticities of the optical field. Moreover, $\boldsymbol{\tau}_{ani}$ tends to drive the nanodumbbell's librations on two degrees of freedom with equilibrium angles $\theta_{equ} = \theta_0$ and $\varphi_{equ} = \frac{\pi}{2}$ respectively.

$\boldsymbol{\tau}_{abs}$ is determined by the total spin angular momenta of absorbed photons per unit time,

$$\boldsymbol{\tau}_{abs} = \mathbf{s}\frac{\sigma_{abs}I_{laser}}{\hbar\omega_{laser}} \quad (9)$$

where $\sigma_{abs}$, $I_{laser}$ and $\omega_{laser}$ stand for the absorption cross-section, the intensity and angular frequency of the trapping laser field respectively, and $\mathbf{s}$ represents the mean spin angular momentum of a single photon. The absorption cross-section $\sigma_{abs}$ is determined by the nanoparticle's geometry, the volume $V$, the complex refraction index $\tilde{n} = n + i\kappa$ and the wavelength $\lambda$. The mean spin angular momentum of elliptically-polarized optical fields $\mathbf{s}$ is derived based on the spin angular momentum operators and the normalized tightly-focused elliptically-polarized optical field. The normalized optical field vector is shown in Equation (10).

$$\mathbf{E}_{inc,normalized} = \frac{1}{\sqrt{|E_x|^2 + |E_y|^2 + |E_z|^2}}\begin{pmatrix} E_x \\ E_y \\ E_z \end{pmatrix} \quad (10)$$

The spin angular momentum operators are shown in Equation (11).

$$\hat{\mathbf{s}}_\mathbf{x} = \begin{pmatrix} 0 & 0 & 0 \\ 0 & 0 & -i \\ 0 & i & 0 \end{pmatrix}\hbar, \hat{\mathbf{s}}_\mathbf{y} = \begin{pmatrix} 0 & 0 & i \\ 0 & 0 & 0 \\ -i & 0 & 0 \end{pmatrix}\hbar, \hat{\mathbf{s}}_\mathbf{z} = \begin{pmatrix} 0 & -i & 0 \\ i & 0 & 0 \\ 0 & 0 & 0 \end{pmatrix}\hbar \quad (11)$$

Therefore, the mean spin angular momentum can be derived as shown in Equation (12).

$$\mathbf{s} = \begin{pmatrix} \mathbf{E}^*_{inc,normalized}\hat{\mathbf{s}}_\mathbf{x}\mathbf{E}_{inc,normalized} \\ \mathbf{E}^*_{inc,normalized}\hat{\mathbf{s}}_\mathbf{y}\mathbf{E}_{inc,normalized} \\ \mathbf{E}^*_{inc,normalized}\hat{\mathbf{s}}_\mathbf{z}\mathbf{E}_{inc,normalized} \end{pmatrix} = \frac{i\hbar}{|E_x|^2 + |E_y|^2 + |E_z|^2}\begin{pmatrix} E_y E_z^* - E_y^* E_z \\ -E_x E_z^* + E_z^* E_x \\ E_x E_y^* - E_x^* E_y \end{pmatrix} \quad (12)$$

Given that longitudinal optical field $|E_z|$ is far lower than transverse optical fields $|E_x|$ and $|E_y|$, $\boldsymbol{\tau}_{abs}$ is dominant along the z axis. The derivations above provide a quantitative explanation why $\tau_{abs,z}$ increases with the ellipticity of the optical field.

## Data availability

Source data are available for this paper. All other data that support the plots within this paper and other findings of this study are available from the corresponding authors upon reasonable request.

## Acknowledgements

We gratefully acknowledge valuable advice from Leiming Zhou at Hefei University of Technology and Xingguang Liu at Harbin Institute of Technology. Cuihong Li is supported by the National Natural Science Foundation of China (NSFC, grant no. 42004154). Chaoxiong He is supported by the NSFC (grant no. 62305305). Huizhu Hu is supported by the Major Scientific Project of Zhejiang Lab (2019MB0AD01). Zhangqi Yin is supported by Beijing Institute of Technology Research Fund Program for Young Scholars. Fu Feng is supported by the NSFC (grant no. 62275167, 92250304).

## Author contributions

Cuihong Li and Zhangqi Yin conceived the idea. Jinchuan Wang, Shaochong Zhu, and Yuanyuan Ma jointly performed the experiments. Qianwen Ying designed the data acquiring system. Chaoxiong He analysed the experimental data, performed numerical calculations, constructed the theoretical model, and wrote the manuscript with the contributions from Cuihong Li, Ying Dong, Zhangqi Yin, Fu Feng and Huizhu Hu. Huizhu Hu, Cuihong Li and Chaoxiong He supported the project.